# A Recount of Sunspot Groups on Staudach's Drawings

Leif Svalgaard[1] (leif@leif.org)

[1] Stanford University, Cypress Hall C13, W.W. Hansen Experimental Physics Laboratory, Stanford University, Stanford, CA 94305, USA

## Abstract:

We have examined the more than 1100 drawings of the solar disk made by the German astronomy amateur Johann Caspar Staudach during 1749-1799 and counted the spots on each image. Using the modern perception of how to group spots into active regions we regrouped the spots as a modern observer would. The resulting number of groups was found to be on average 25% higher than the first count of groups performed by Wolf in 1857, and used by Hoyt and Schatten in their construction of the Group Sunspot Number. Compared to other observers at the time, Staudach's drawings have a very low average number, ~2, of spots per group, possibly indicating an inferior telescope likely suffering from spherical and chromatic aberration as would typical of amateur telescopes of the day. We have initiated an ongoing project aiming at observing sunspots with antique telescopes having similar defects in order to determine the factor necessary to bring the Staudach observations onto a modern scale.

# 1. Introduction

In a short book by Lorenz Woeckel "Die Sonne und ihre Flecken" (Woeckel, 1846) there is a reference to a Nürnberger: Johann Caspar Staudach (1731-1799? - for the preferred spelling see Spörer, 1887) who in the years 1749 to 1799 made numerous, although not quite systematic, observations of sunspots using a helioscope (this method for observing sunspots was first used by Galileo's student Benedetto Castelli and involves projecting a telescopic image of the sun onto a white sheet of paper in a darkened room). Woeckel added that the Staudach observations confirmed Schwabe's conclusion about a 10-year period in sunspot occurrence "as was readily seen from the summary table" (given by Woeckel). Rudolf Wolf didn't quite agree that the table was all that clear (disagreeing in parts with this own) and sought to clarify the situation by getting access to the original Staudach observations consisting of about 1000 drawings of sunspots on the solar disk. He was fortunate that mediation by Culmann and Bauernfeind resulted in obtaining in short order the original publication by the "trusting kindness" of its current owner Mr. G. Eichhorn in Nürnberg. Wolf later returned the material. Gustaf Spörer (1887) writes in a letter to Wolf that the "Staudach manuscript has recently, due to lucky chance, been saved from destruction and has now landed in my hands". Spörer alerts Wolf to a couple of misprints in Wolf's tabulation, but does not otherwise comment on the data.

# 2. Staudach's Telescope

The drawings are today stored in the library of the Astrophysikalisches Institut Potsdam, Germany, and are in very good condition. Arlt (2008) has recently photographed the drawings. The images in JEPG format produced by the camera are considered of sufficient quality for counting sunspots. Arlt also draws some inference about the telescope used by Staudach. In the material there is a single mention of a telescope (18 February 1775: "when I turned round with my 3-foot sky tube…") hence we may assume that the focal length of the telescope was 3 feet. Achromatic telescopes with a focal length of 92 cm were manufactured by John and Peter Dollond from the late 1750s. With such a telescope, however, the distinction between umbra and penumbra should have been possible, and the Wilson effect should have been visible. Both were not noted by Staudach and were not clearly present. Staudach started with a different style of drawing on 1768 December 2. Arlt (2008) notes that large spots are often surrounded by many small dots representing either the penumbra or smaller spots. Perhaps Staudach obtained an improved telescope in 1768 causing him to show indications of a penumbra, or he had the chance to see sunspots with another telescope, or had heard about more precise drawings of sunspots, that he started indicating structure in 1768, but more from knowledge than from observation (Arlt 2008), but see Section 4.

An average telescope used by an amateur at the time probably suffered from fairly strong spherical aberration. Because of a couple of mirrored solar-eclipse drawings, Arlt (2008) suggests that Staudach was using a Keplerian refractor with a non-achromatic objective and that he most likely missed all the A and B spot groups (according to the Waldmeier (1938) classification). Such groups make up 30-50% of all groups seen today. To convert a group count without A and B groups to a full count of groups of all classes, one must thus multiply by 1.65, which incidentally is the same factor it takes to reduce the group count obtained by Wolf using his small, but superb, 2½-foot Fraunhofer refractor to the count by his successor Wolfer, using the 4-foot norm-telescope (Svalgaard and Schatten, 2015). Taking into account that Staudach's telescope likely suffered from both spherical and chromatic aberration, the actual factor is likely to be significantly larger. I have initiated an observing program in collaboration with the Antique Telescope Society (http://webari.com/oldscope/) with several amateur observers using old telescopes from the 18th century with characteristics believed to be similar to Staudach's telescope(s) with the goal of assessing a likely value of the reduction factor. The results will be published in a later paper.

## 3. Counting Groups on Staudach's Drawings

Wolf (1857) wrote about the Staudach drawings that "even if the groups could not always be distinguished with certainty from each other, I nevertheless decided to count not only the spots, but also the number of groups as best as possible". Wolf's group counts were apparently used without changes (apart from a few clerical errors) in the construction of the Group Sunspot Number by Hoyt and Schatten (1998, hereafter HS). Figure 1 (from Clette et al. (2014)) shows typical examples of the drawings. Wolf (and thus HS) reported only one group on both days. A modern observer would interpret the spot grouping differently. I see at least three groups, marked by the ovals. Asking several active observers (among them Sergio Cortesi, the reference observer in Locarno) confirms my estimate, and prompted an effort to re-count the groups on Staudach's drawings from a modern perspective.

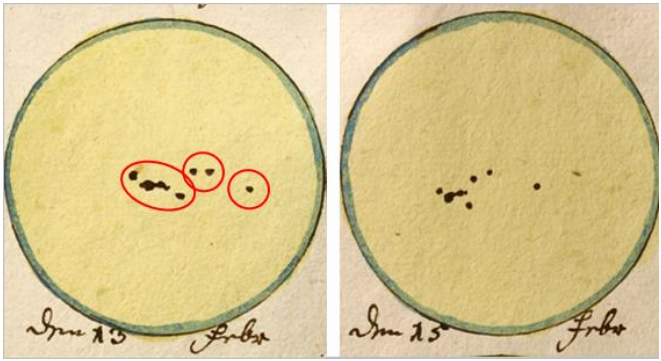

Figure 1: Staudach's drawings for the 13th (left) and 15th (right) of February, 1760. On both these days Wolf (and HS) reported only a single group; a modern observer would count at least three groups on those days as indicated by the red ovals (*e.g.* S. Cortesi, Pers. Comm.). Image Courtesy R. Arlt.

Lower panel: The group count from HS's database for February 1760 (left), compared to counts from Wolf's tabulation (1857) in the format **groups.spots** (right). The number of spots (7 and 8) reported by Wolf match well what a modern observer would find on the drawings.

| Day | Jan | Feb |
|-----|-----|-----|
| 11 | -99 | 3 |
| 12 | -99 | 1 |
| 13 | -99 | ①  |
| 14 | -99 | -99 |
| 15 | -99 | ①  |
| 16 | -99 | -99 |
| 17 | -99 | -99 |
| 18 | -99 | 2 |

| II | 11 | 3.5 |
|----|----|-----|
| — | 12 | 1.7 |
| — | 13 | 1.7 |
| — | 15 | 1.8 |
| — | 18 | 2.5 |

We do not, of course, know how Wolf grouped the spots into groups, but it is generally possible to find a plausible grouping matching Wolf's count, for example as shown in Figure 2. Even if our guess about how Wolf grouped the spots were wrong that would not change *our* re-assessment or re-count of the groups.

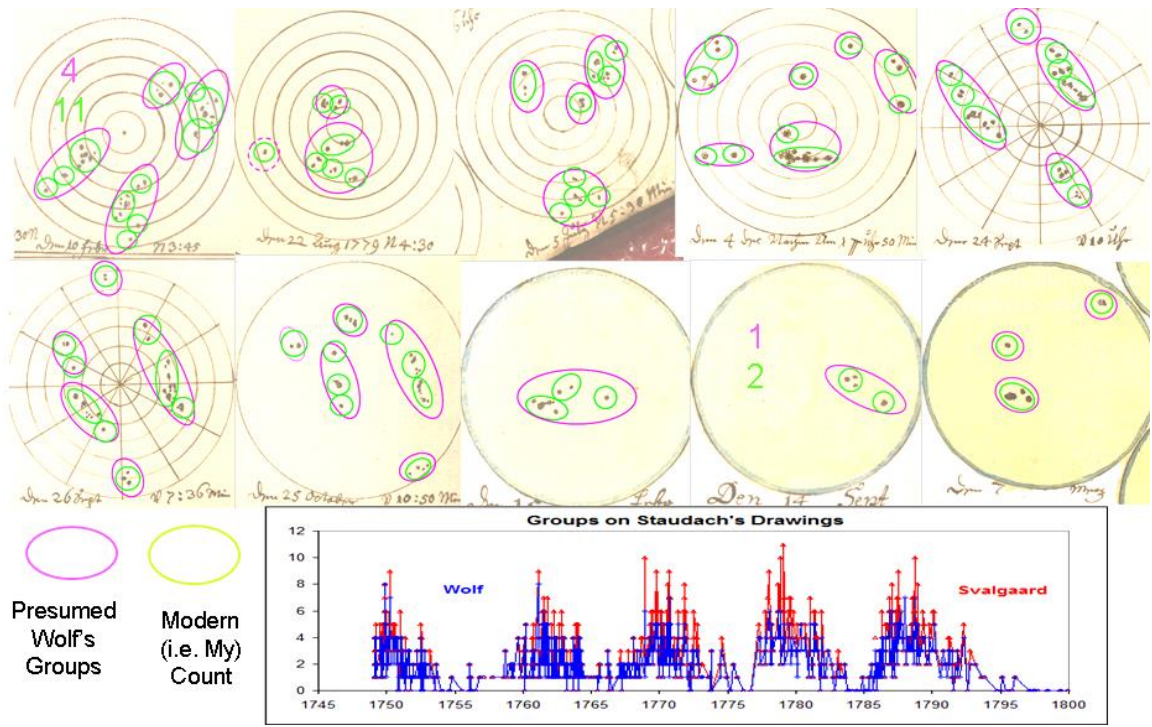

Figure 2: Several examples of grouping of sunspots. Wolf's presumed groupings are shown as pink ovals, while my determinations of the groups are shown as green ovals. The lower panel shows daily values of Wolf's (blue) and my counts (red), both dominated, of course, by the largest values.

Figure 3 shows that the difference between my count and Wolf's count does not depend on the number of groups. On average, my count is 25% larger than Wolf's, regardless of the amount of solar activity, as measured by the number of groups. On the other hand, as one would expect, the number of spots counted is almost the same.

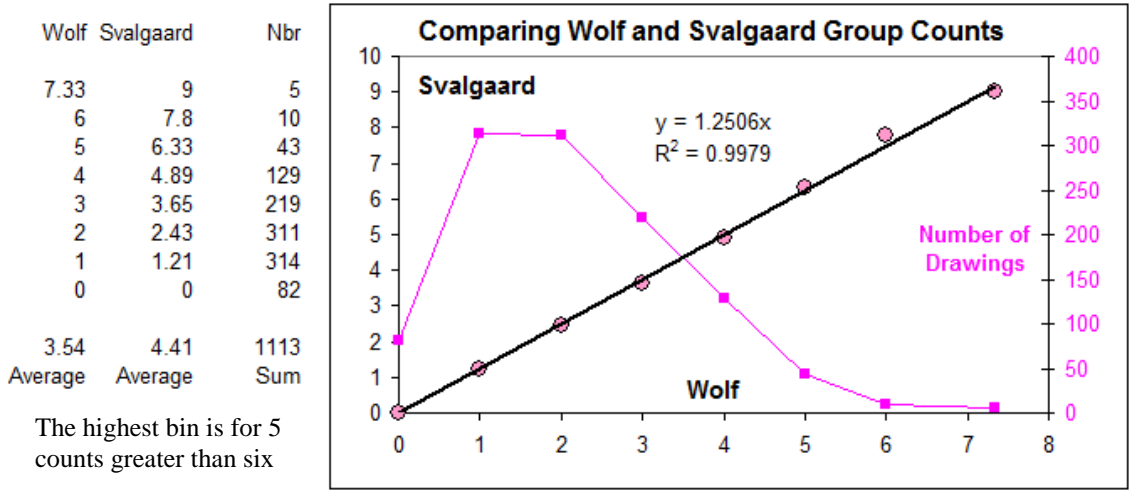

| Wolf | Svalgaard | Nbr |
|---|---|---|
| 7.33 | 9 | 5 |
| 6 | 7.8 | 10 |
| 5 | 6.33 | 43 |
| 4 | 4.89 | 129 |
| 3 | 3.65 | 219 |
| 2 | 2.43 | 311 |
| 1 | 1.21 | 314 |
| 0 | 0 | 82 |
| | | |
| 3.54 | 4.41 | 1113 |
| Average | Average | Sum |

The highest bin is for 5 counts greater than six

Figure 3: The number of groups counted by Wolf compared to my count. The pink curve shows the number of drawings for each bin of Wolf's count.

In Clette at al. (2014) and Svalgaard and Schatten (2015) we introduced the "backbone" method for reconstructing solar activity in the past: find a primary observer for a certain (long) interval and normalize all other observers individually to the primary based on overlap with only the primary (minimizing accumulation of errors). The selection of the primary observer should be based both on the length of the observational series (as long as possible) and on the perceived "quality" of the observations such as regularity of observing, suitable telescope, and lack of obvious problems. Each backbone is "floating" (i.e. tied to the primary observer) and must be harmonized with other backbones to form a final composite. Figure 4 compares the Wolf and Svalgaard group counts with the floating Staudach backbone, the latter based on an average of 3.4 observers per year.

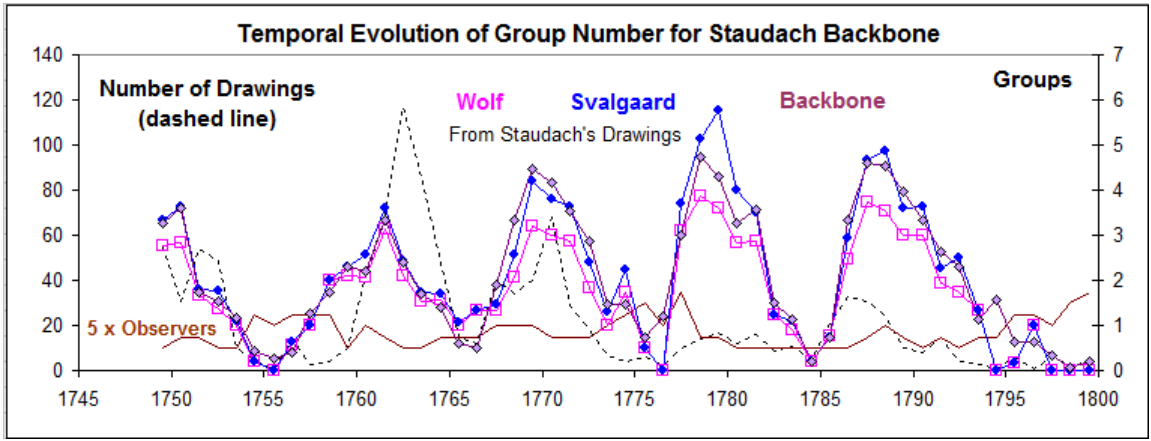

Figure 4: Yearly averages of Wolf's count (pink squares), my count (blue diamonds) and the composite (but floating) Staudach backbone using all available observers (purple diamonds). The left-hand scale is for the number of drawings per year (dashed curve) and for five times the number of observers (average 3.4) used for this part of the backbone (brown curve).

## 4. Spots per Group

The weaker the telescope is, the fewer spots per group an observer will report (or draw). In addition, the counting method matters, e.g. whether the smallest spots on the limit of visibility are counted or not. This is vividly illustrated in Figure 5 which shows the average number of spots per group reported by the Zürich observers 1849-1901 using three different telescopes and two different counting methods:

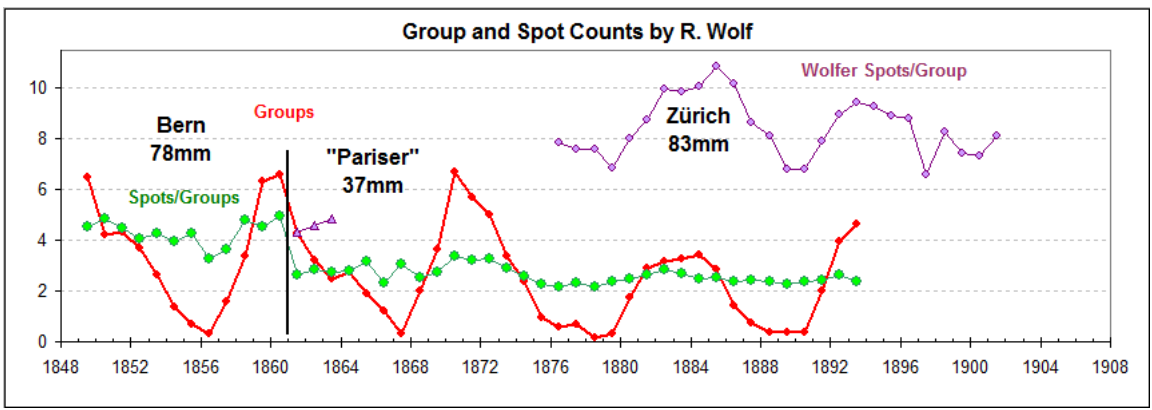

Figure 5: Yearly averages of the number of spots per group reported by Wolf (green circles) and by Wolfer (purple circles).

Before 1861 Wolf used a 78 mm aperture refractor in Bern (the whereabouts today of the telescope is unknown) and also for three years after that on occasional visits to Bern (purple triangles). The number of spots per group for those observations was 4.5. From 1861 and until his death in 1893 Wolf used smaller (37-40 mm aperture) handheld telescopes and the number of spots per group was an accordingly smaller 2.6. Wolf originally did not count the smallest spots on the limit of visibility in order to be compatible with Schwabe's counting method. On the other hand, his successor Wolfer insisted on counting all spots that he could see and so (even with a telescope of almost the same aperture as the one in Bern) reported significantly more spots per group than Wolf with the similar telescope, namely 8.4 compared to 4.5.

Inherent in the definition of the relative sunspot number, R, is the stipulation (when Wolf chose 10 as the weight for Groups in his definition of the Relative Sunspot Number, he remarked that he could as well have chosen 9 or 11, but that 10 was certainly "more convenient") that the number of spots per group is constant, R = 10×Groups + Spots, but this assumption is only approximately true, there is both a sunspot cycle variation and possibly longer period secular variations, Figure 6:

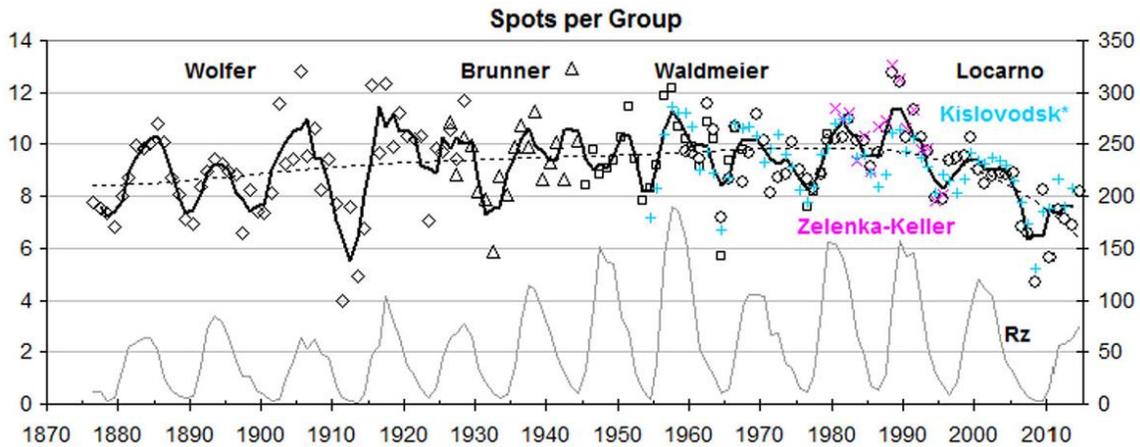

Figure 6: The ratio between the number of single spots and the number of groups as recorded by the Zürich observers for each year of observation. The full curve is a 3-year running mean. There is clear solar cycle dependence (Sunspot Number Version 1 shown at bottom) with more spots per group at higher solar activity. From Clette et al. (2014).

It is not clear, yet, how to separate the solar variations of the spot to group ratio from the combined effects of telescope quality and counting method, but we can at least determine the spot/group ratio for Staudach using my re-counted group numbers, Figure 7. For the whole record the ratio is 2.02±0.07, with most of the variance due to the solar cycle variation of the ratio, also discernible in Figure 7. The ratio is much lower than Wolf's, probably attesting to the low-quality telescope and the crude drawings. A vertical line (at the end of 1768) divides the periods of different drawing styles. It is doubtful if the difference in the ratio before and after the division is significant, tacking into account that the levels of solar activity were also different.

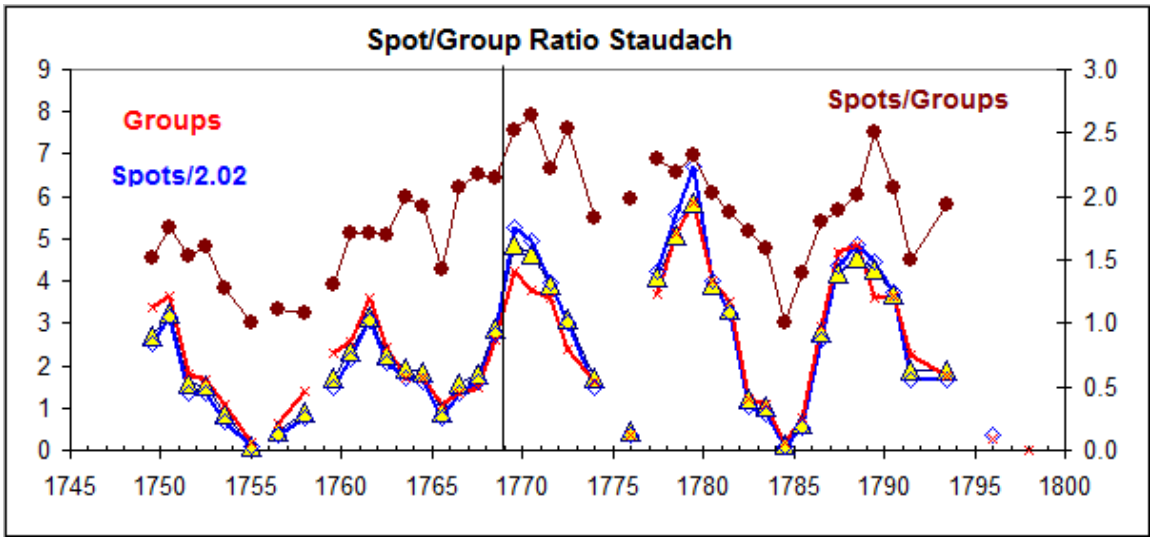

Figure 7: The number of spots per group (brown symbols, right-hand scale, my re-count) for Staudach's drawings. On average there are 2.02 spots per group, so the number of spots divided by 2.02 (blue symbols) should to first order vary like the number of groups (red symbols). A second-order fit (yellow triangles; $R^2 = 0.943$) is a better match. The vertical line at the end of 1768 marks the division between the two drawing styles.

## 5. Comparisons with Other Observers

Although the observations by other observers are scattered and often unsystematic it is of notice that their group counts are generally higher than those of Staudach, possibly indicative of his poorer telescope. Figure 8 shows the yearly average group counts by all observers active during the Staudach era.

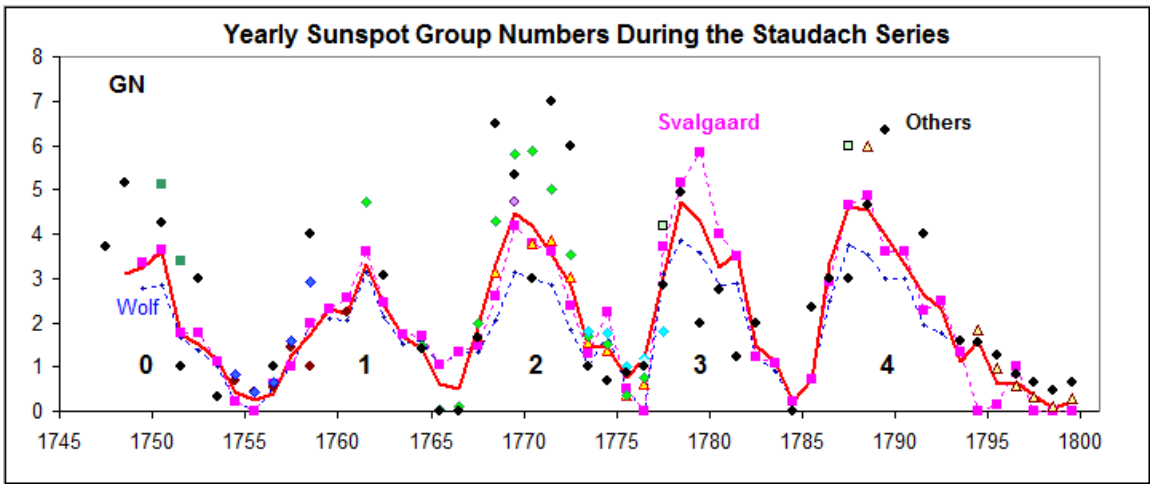

Figure 8: Average number of sunspot groups per year for different observers. Broken blue line: Wolf's count of Staudach (1857); broken pink line with squares: Svalgaard's count of Staudach; solid red curve (this article): the floating Staudach backbone (Svalgaard and Schatten, 2015); yellow triangles with red border: Hoyt and Schatten's count of Horrebow (1995); green

diamonds: d'Arrest's count of Horrebow (Wolf, 1873); purple diamond: Wolf's count of Horrebow (Wolf, 1865); brown diamonds: Wolf's count of Zucconi (Wolf, 1857); green squares: Wagner's count of Hagen (Wolf, 1859); blue diamonds: Kayser's count of Schubert (Wolf, 1870); light blue diamonds: Wolf's count of Mallet (Wolf, 1858); medium blue diamonds and open squares: Hoyt and Schatten's count of Lievog and Bugge (1995); yellow triangles with brown border: Wolf's count of Flaugergues (1861); black diamonds: average of Hoyt and Schatten's counts of all other (~2.5 per year) observers (1998).

The yearly values are also given in Table 1. Daily values 1749-1799 (which also include the number of spots) are available as supplemental data and also at the author's website http://www.leif.org/research, item 2260.

## 6. Conclusion

We have examined the drawings of the solar disk made by Staudach during 1749-1799 and counted the spots on each image. Using the modern perception of how to group spots into active regions we regrouped the spots as a modern observer would. The resulting number of groups was found to be on average 25% higher than the first count of groups performed by Wolf, and used by Hoyt and Schatten. Compared with other observers at the time, Staudach's drawings have a very low average number, ~2, of spots per group, possible indicating an inferior telescope likely suffering from spherical and chromatic aberration. We have initiated a project aiming at observing sunspots with antique telescopes having similar defects in order to determine the factor necessary to bring the Staudach observations onto a modern scale. The results of this effort will be reported in a coming article.

Table 1: Yearly average group counts for the data plotted in Figure 8. The 'which' and 'who' rows correspond to the observers and 'reducers' detailed in the caption for the Figure.

| Which: | St | St | Ho | Ho | Ho | Zu | Ha | Sc | Ma | LB | Fl | AO |
| Who: | RW | LS | DA | HS | RW | RW | RW | KA | RW | HS | RW | HS |
| 1747.5 | | | | | | | | | | | | 3.70 |
| 1748.5 | | | | | | | | | | | | 5.17 |
| 1749.5 | 2.76 | 3.34 | | | | | | | | | | |
| 1750.5 | 2.83 | 3.64 | | | | | 5.10 | | | | | 4.25 |
| 1751.5 | 1.66 | 1.78 | | | | | 3.39 | | | | | 1.00 |
| 1752.5 | 1.38 | 1.75 | | | | | | | | | | 3.00 |
| 1753.5 | 1.00 | 1.10 | | | | | | | | | | 0.33 |
| 1754.5 | 0.20 | 0.20 | | | | 0.68 | | 0.83 | | | | |
| 1755.5 | 0.00 | 0.00 | | | | 0.43 | | 0.44 | | | | |
| 1756.5 | 0.53 | 0.61 | | | | 0.53 | | 0.64 | | | | 1.00 |
| 1757.5 | 1.00 | 1.00 | | | | 1.45 | | 1.58 | | | | |
| 1758.5 | 2.00 | 2.00 | | | | 1.00 | | 2.93 | | | | 4.00 |
| 1759.5 | 2.10 | 2.30 | | | | | | | | | | |
| 1760.5 | 2.07 | 2.57 | | | | 2.25 | | | | | | |
| 1761.5 | 3.14 | 3.60 | | 4.71 | | | | | | | | |
| 1762.5 | 2.11 | 2.44 | | | | | | | | | | 3.07 |

|  |  |  |  |  |  |  |  |  |  |  |  |
|---|---|---|---|---|---|---|---|---|---|---|---|
| 1763.5 | 1.53 | 1.73 |  |  |  |  |  |  |  |  |  |
| 1764.5 | 1.58 | 1.70 |  | 1.47 |  |  |  |  |  |  | 1.40 |
| 1765.5 | 1.00 | 1.06 |  | 0.02 |  |  |  |  |  |  | 0.00 |
| 1766.5 | 1.33 | 1.33 |  | 0.11 |  |  |  |  |  |  | 0.00 |
| 1767.5 | 1.35 | 1.47 | 1.68 | 2.00 |  |  |  |  |  |  | 1.65 |
| 1768.5 | 2.05 | 2.58 | 3.15 | 4.30 |  |  |  |  |  |  | 6.50 |
| 1769.5 | 3.13 | 4.18 |  | 5.81 | 4.71 |  |  |  |  |  | 5.33 |
| 1770.5 | 2.98 | 3.79 | 3.80 | 5.89 |  |  |  |  |  |  | 3.00 |
| 1771.5 | 2.86 | 3.62 | 3.84 | 5.02 |  |  |  |  |  |  | 7.00 |
| 1772.5 | 1.83 | 2.38 | 3.02 | 3.53 |  |  |  |  |  |  | 6.00 |
| 1773.5 | 1.00 | 1.28 | 1.54 | 1.71 |  |  |  |  | 1.80 |  | 1.00 |
| 1774.5 | 1.75 | 2.25 | 1.36 | 1.51 |  |  |  |  | 1.75 |  | 0.70 |
| 1775.5 | 0.50 | 0.50 | 0.35 | 0.35 |  |  |  |  | 1.00 |  | 0.87 |
| 1776.5 | 0.00 | 0.00 | 0.60 | 0.76 |  |  |  |  | 1.20 | 1.10 | 1.00 |
| 1777.5 | 3.10 | 3.70 |  |  |  |  |  |  | 1.80 | *3.51* | 2.86 |
| 1778.5 | 3.85 | 5.14 |  |  |  |  |  |  |  |  | 4.95 |
| 1779.5 | 3.58 | 5.82 |  |  |  |  |  |  |  |  | 2.00 |
| 1780.5 | 2.83 | 4.00 |  |  |  |  |  |  |  |  | 2.75 |
| 1781.5 | 2.87 | 3.50 |  |  |  |  |  |  |  |  | 1.22 |
| 1782.5 | 1.22 | 1.22 |  |  |  |  |  |  |  |  | 2.00 |
| 1783.5 | 0.90 | 1.09 |  |  |  |  |  |  |  |  |  |
| 1784.5 | 0.20 | 0.20 |  |  |  |  |  |  |  |  | 0.00 |
| 1785.5 | 0.76 | 0.72 |  |  |  |  |  |  |  |  | 2.33 |
| 1786.5 | 2.48 | 2.93 |  |  |  |  |  |  |  |  | 3.00 |
| 1787.5 | 3.74 | 4.65 |  |  |  |  |  |  | 6.00 |  | 3.00 |
| 1788.5 | 3.54 | 4.86 |  |  |  |  |  |  |  | 6.00 | 4.67 |
| 1789.5 | 3.00 | 3.60 |  |  |  |  |  |  |  |  | 6.33 |
| 1790.5 | 3.00 | 3.62 |  |  |  |  |  |  |  |  |  |
| 1791.5 | 1.93 | 2.26 |  |  |  |  |  |  |  |  | 4.00 |
| 1792.5 | 1.75 | 2.50 |  |  |  |  |  |  |  |  |  |
| 1793.5 | 1.33 | 1.33 |  |  |  |  |  |  |  |  | 1.57 |
| 1794.5 | 0.00 | 0.00 |  |  |  |  |  |  |  | 1.82 | 1.54 |
| 1795.5 | 0.16 | 0.16 |  |  |  |  |  |  |  | 0.97 | 1.25 |
| 1796.5 | 1.00 | 1.00 |  |  |  |  |  |  |  | 0.59 | 0.83 |
| 1797.5 | 0.00 | 0.00 |  |  |  |  |  |  |  | 0.32 | 0.63 |
| 1798.5 | 0.00 | 0.00 |  |  |  |  |  |  |  | 0.11 | 0.46 |
| 1799.5 | 0.00 | 0.00 |  |  |  |  |  |  |  | 0.28 | 0.66 |

## Acknowledgements

We thank Rainer Arlt for making available the photographic images of the Staudach drawings. LS thanks Stanford University for continuing support.

## References

Arlt, R.: 2008, Digitization of Sunspot Drawings by Staudacher in 1749 – 1796, *Solar Phys.* **247**, 399, doi 10.1007/s11207-007-9113-4

Clette, F., Svalgaard L., Vaquero, J. M., and Cliver, E. W.: 2014, Revisiting the sunspot number: A 400-year perspective on the solar cycle, *Space Sci. Rev.* **16**, 35, doi:10.1007/s11214-014-0074-2

Hoyt, D. V., Schatten, K. H.: 1995, A new interpretation of Christian Horrebow's sunspot observations from 1761 to 1777, *Solar. Phys.* **160**, 387, doi:10.1007/BF00732817

Hoyt, D. V., Schatten, K. H.: 1998, Group sunspot numbers: a new solar activity reconstruction, *Solar. Phys.* **181**, 491, doi:10.1023/A:1005056326158

Spörer, G.: 1887, in letter to Wolf in Mittheilungen über die Sonnenflecken, **LXIX**, item 547

Svalgaard, L. and Schatten, K. H.: 2015, Reconstruction of the Sunspot Group Number: the Backbone Method, *Solar Phys.* **XXX**, yyy, doi 10.1007/s11ddddddddddd

Waldmeier, M.: 1938, Chromosphärische Eruptionen, *Z. Astrophys.* **16**, 276

Woeckel, L.: 1846, *Die Sonne und ihre Flecken*, Campe, Nürnberg, 31pp
https://download.digitale-sammlungen.de/pdf/1445541475bsb10049284.pdf

Wolf, R.: 1857, Mittheilungen über die Sonnenflecken, **IV**, 52

Wolf, R.: 1858, Mittheilungen über die Sonnenflecken, **VII**, 160

Wolf, R.: 1859, Mittheilungen über die Sonnenflecken, **IX**, 208

Wolf, R.: 1861, Mittheilungen über die Sonnenflecken, **XIII**, 83

Wolf, R.: 1865, Mittheilungen über die Sonnenflecken, **XIX**, 263

Wolf, R.: 1870, Mittheilungen über die Sonnenflecken, **XXVII**, 239

Wolf, R.: 1873, Mittheilungen über die Sonnenflecken, **XXXIII**, 120